\documentclass[12pt,]{article}
\usepackage{amsmath}
\usepackage{amsfonts} \usepackage{amssymb}
\newtheorem{theorem}{Theorem}[section]

\newtheorem{proposition}[theorem]{Proposition}

\numberwithin{equation}{section}

\begin{document}      


\title{On the AdS stability problem.}

\author {Helmut Friedrich\\
Max-Planck-Institut f\"ur Gravitationsphysik\\
Am M\"uhlenberg 1\\ 14476 Golm, Germany }

\maketitle                
 
\begin{abstract}

We discuss the notion of stability and the choice of boundary conditions for AdS-type space-times and point out difficulties  in the construction of Cauchy data which arise   if reflective boundary conditions are imposed.

\end{abstract}

{\footnotesize

\section{Introduction}

P. Bizo\'n and A. Rostworowski recently presented a study of
the  stability of anti-de Sitter  space (\cite{bizon:rostworowski:2011}) which raises some extremely interesting questions concerning
solutions to Einstein's field equations
\begin{equation}
\label{I-Einst-equ}
\hat{R}_{\mu\nu} - \frac{1}{2}\,\hat{R}\,\hat{g}_{\mu\nu} + \lambda\,\hat{g}_{\mu\nu}
= \kappa\,\hat{T}_{\mu\nu},
\end{equation}
with cosmological constant $\lambda < 0$ that are subject to conditions on the boundary ${\cal J}$ at space-like and null infinity.
They analyse  the spherically symmetric Einstein-massless-scalar field 
system with homogeneous Dirichlet asymptotics and Gaussian type initial data 
and observe the formation of trapped  surfaces for (numerically)  arbitrarily small  initial data. 
They perform a perturbative analysis, which points into the same direction but also exhibits  
{\it small one-mode initial data} which develop into globally smooth solutions.
Further, they supply  numerical evidence that the development of trapped surfaces results 
from an energy transfer from low to high frequency modes. Of the follow-up work
(\cite{buchel-lehner-liebling:2012},
\cite{buchel-lehner-liebling:2013},
\cite{dias:horowitz:santos:2012}, \cite{dias:horowitz:marolf:santos:2012},
\cite{ishibashi:maeda:2012}) we only  mention the observation in
\cite{buchel-lehner-liebling:2013}
that also solutions arising  from data sufficiently close to the
small one-mode initial data  exist (numerically) for all time.

Their  results led P. Bizo\'n and A. Rostworowski  to conjecture: {\it AdS is unstable against the formation of black holes for a large class of arbitrarily small perturbations}. This conjecture has been formulated in the context of a particular model but  it may easily be misunderstood as a statement applying to more general situations.
The purpose of the present note is to point out that such a conclusion may possibly be too strong if the class  of  competing perturbations is reasonably large and it should rather be replaced by: {\it AdS with reflecting boundary conditions is unstable against the formation of black holes for a large class of arbitrarily small perturbations}. This is still a very interesting conjecture. Understanding in detail how the  focussing property inherent in the non-linear equations combines with the, in principle unlimited in time, refocussing effected at the reflecting boundary should give important insights into the evolution process. 

By many workers in the field reflecting boundary conditions are so naturally associated with  
AdS-type solutions  that they are sometimes simply referred to as `the AdS boundary conditions'. 
They are clearly convenient because they  provide clean initial boundary value problems which exclude any information entering or leaking out of the system. But as a consequence, these systems cannot  interact with an ambient cosmos and thus certainly  do not represent observable objects as suggested by some of the names  given to them in the literature. Their astrophysical  interest thus remains unclear. Reflecting boundary conditions are very special and as long as it is not clear what kind of objects the solutions represent from the point of view of physics (and why a negative cosmological constant should be introduced in the first place) it does not appear reasonable to exclude all others possibilities.

If more general boundary conditions are admitted there may still be arbitrarily small perturbation leading to trapped surfaces but the class of these perturbations may be much smaller relative to the complete set of admitted perturbations
because radiation can enter and leave the system through the boundary ${\cal J}$ at space-like and null infinity over an unlimited length of (conformal) time. It is here where the problem of stability becomes for AdS-type space-times much more difficult than in the case of de Sitter-type or Minkowski-type solutions. In the latter cases the conformal boundary consists of two components:  ${\cal J}^+$ through which radiation  can leave the system and ${\cal J}^-$ across which radiation can enter the system and conveniently be controlled, at least in principle, in its  size  and form. The possibility of 
such a clear distinction is not obvious in AdS-type space-times and whether any kind of  analogue to this can be established in terms of boundary data/conditions on ${\cal J}$ is part of the more general stability problem.

The latter problem will not be considered in this article. Instead, we shall revisit a known existence result for AdS-type space-times which exhibits the full freedom to prescribe boundary data on ${\cal J}$ and discuss various aspects pertaining to the given problem in some detail. In particular,  reflecting boundary conditions will be reconsidered from this more general perspective. Here comes into play a specific feature of the initial boundary value problems for Einstein's field equations: If conditions are imposed  on the boundary data, the clean separation between the evolution problem and the analysis of the constraints on the initial space-like slice, usual in the standard Cauchy problem, cannot be maintained any longer. As a consequence, reflective boundary conditions lead  to a type of problem for the Cauchy data on the initial space-like slice which has not been discussed so far. Besides the standard constraints and the hyperboloidal fall-off behaviour required at space-like infinity the data are constrained by a sequence of additional  conditions at space-like infinity. These encode the requirements of reflective boundary conditions in the structure of the Cauchy data. Whether this may give a way to decide whether the {\it class of arbitrarily small perturbations which  lead to the formation of black holes} is relatively increased by imposing reflective boundary conditions remains to be seen.

\section{AdS-type solutions}

In four space-time dimensions anti-de Sitter covering space, short AdS, is given by
\[
\hat{M} = \mathbb{R} \times \mathbb{R}^{3}, \quad 
\hat{g} =  - \cosh^2 r\,dt^2 + dr^2 + \sinh^2 r\,h_{\mathbb{S}^{2}},
\]
where $r \ge 0$ denotes the standard radial coordinate on $ \mathbb{R}^{3}$ and 
$h_{\mathbb{S}^{2}}$ the standard round metric on $ \mathbb{S}^{2}$.
It is a static solution to 
\[
R_{\mu \nu}[\hat{h}] = \lambda\,\hat{g}_{\mu \nu} \quad \mbox{with} \quad
\lambda = - 3.
\quad \quad
\]
On the space-like slices $t = const.$ it induces a metric of constant negative curvature which can be conformally compactified and the clearest picture of its global features is obtained by 
performing the related conformal extension of AdS. Combining the coordinate transformation 
$r \rightarrow \rho = 2\,\arctan(e^r) - \frac{\pi}{2}$ with the rescaling by the conformal factor
$\Omega = \frac{1}{\cosh r} = \cos \rho$, both depending only on the spatial coordinates,
gives the conformal representation
\[
g = \Omega^2\,\hat{g} = - dt^2 + d\rho^2 + \sin^2\rho\,\,h_{\mathbb{S}^{2}},\quad 
t \in \mathbb{R}, \quad 0 \le \rho < \frac{\pi}{2},
\]
of AdS which induces on the slices $t = const.$
the standard round metric on $\mathbb{S}^{3}$. The metric $g$ extends smoothly as $\rho \rightarrow \frac{\pi}{2}$ and then lives on the manifold
$M = \mathbb{R} \times  \frac{\mathbb{S}^{3}}{2}$
with $g$-time-like conformal boundary
${\cal J} = \{\rho = \frac{\pi}{2}\} \sim  \mathbb{R} \times \mathbb{S}^{2}$. The boundary points can be understood  as endpoints of the space-like and null geodesics and ${\cal J}$ as representing space-like and null infinity for AdS.

In the following a solution of Einstein's field equations with negative cosmological constant which admits in a similar way a smooth conformal extension that  adds a time-like hypersurface ${\cal J}$ representing space-like and null infinity will be referred to as {\it  AdS-type space-time.}
(cf. \cite{penrose:scri} for details).
There may be notions of {\it asymptotically AdS space-times} involving boundaries at space-like and null infinity of lower smoothness 
but for the sake of our discussion `AdS-type' as above is convenient. What will be said in the following  may apply to more general situations but working out the consequences will also be more difficult. 

Two features of the global  causal/conformal structure of AdS make  global 
problems involving AdS-type solutions 
 quite different from those for de Sitter-type or Minkowski-type solutions.  The first basic observation, emphasized  almost everywhere, is that such space-times are not globally hyperbolic; a time-like curve can always decide to escape through the conformal boundary before hitting a given achronal hypersurface. The second observation, almost never mentioned though of equal importance in the stability problem, is  that AdS
 does not admit a finite conformal representation of past/future time-like infinity {\it which is also smooth}. In this sense AdS is always  infinite in time, also in conformal time.
(We are not referring here to AdS-type solutions because no completeness in time-like directions has been required for them so far).

\section{Well-posed initial boundary value problems for AdS-type solutions }

Expanding an AdS-type vacuum solution  in the conformal setting formally  in terms of a spatial coordinate $r$ which vanishes on ${\cal J}$ 
 gives a Taylor expansion in powers of $r$ if the space-time dimension is even and 
 a `poly-homogeneous expansion' in terms of powers of  $r$ and $\log r$  if the space-time dimension is odd (cf. \cite{fefferman:graham:1985}, \cite{graham:hirachi:2005}). If the data prescribed on ${\cal J}$ are real analytic and the formal expansion is Taylor this procedure yields real analytic AdS-type solutions in some neighbourhood of 
 ${\cal J}$. While they may be useful in some contexts, we shall not consider such solutions. Analyticity is not a desirable 
 assumption for our purpose,  Cauchy problems for hyperbolic equations with data on time-like hypersurfaces are known to be not well-posed in the category of  $C^{\infty}$ functions, and, in particular, 
it is not clear under which conditions on the data such solutions will extend so as to define a smooth interior or possibly another component of the conformal boundary.

The natural problem here is the initial boundary value problem with boundary data prescribed on ${\cal J}$  and Cauchy data on a space-like slice that extends to the boundary.
The basic question then is: How must boundary conditions/data be prescribed so as  to obtain 
well-posed  initial boundary value problems for Einstein's equations, possibly coupled to some matter fields, that produce AdS-type solutions ? 
There exists quite some literature about field equations on AdS- or asymptotically AdS-backgrounds
(\cite{avis:isham:storey:1978}, \cite{bachelot:2008},  \cite{breitenlohner:freedman:1982},
\cite{holzegel:2012}, \cite{ishibashi:wald:2004}, \cite{vasy:2010}, \cite{warnick:2012})
in which ill- and well-posed initial boundary value problems for different matter fields are discussed.
The  boundary conditions and data which can be given on ${\cal J}$ and the behaviour of the solutions 
near ${\cal J}$ will clearly depend on the nature of the test field equations and their behaviour under conformal rescalings.  For conformally covariant field equations the boundary ${\cal J}$ is as 
good as any other time-like hypersurface and offers the same freedom to prescribe boundary data. The boundary analysis can be much more difficult, however,  for `conformally ill-behaved'  equations. 
The huge spectrum of possibilities has hardly been explored so far and very little is known 
 for Einstein's field equations coupled to matter fields if no symmetries are assumed.
We shall thus consider Einstein's field equations with $\lambda < 0$ in  the vacuum case 
$\hat{T}_{\mu \nu} = 0$. 
The following results in four space-time dimensions 
have been obtained by working  in the conformal picture and using the {\it conformal field equations} (\cite{friedrich:AdS}). In the following we shall only discuss some aspects of it; for details we refer the reader to 
the original article.

\begin{theorem} 
\label{ads-exists}
Suppose $\lambda < 0$ and $(\hat{S}, \hat{h}_{ab}, \hat{\chi}_{ab})$ is a smooth Cauchy  data set for $Ric[\hat{g}] = \lambda\,\hat{g}$ so that  $\hat{S}$ is an open, orientable,  $3$-manifold and
$(\hat{S}, \hat{h}_{ab})$ is a complete Riemannian space. Let these data  admit a smooth conformal completion 
\[
\hat{S} \rightarrow S = \hat{S} \cup \Sigma, \quad
\hat{h}_{ab} \rightarrow h_{ab} = \Omega^2\,\hat{h}_{ab}, \quad
\hat{\chi}_{ab} \rightarrow \chi_{ab} = \Omega\,\hat{\chi}_{ab},
\]
so that $(S, h_{ab})$ is a Riemannian space with compact boundary $\Sigma =  \partial S$, 
$\Omega$ a defining function of $\Sigma$,  and  
$W^{\mu}\,_{\nu \lambda \rho} = \Omega^{-1}\,C^{\mu}\,_{\nu \lambda \rho}$ 
extends smoothly to $\Sigma$ on $S$ where $C^{\mu}\,_{\nu \lambda \rho}$ denotes then conformal Weyl tensor determined by the metric $h_{ab}$ and the second fundamental form $\chi_{ab}$.

Consider the boundary ${\cal J} = \mathbb{R} \times \partial S$ of $M = \mathbb{R} \times S$ 
and  identify 
$S$ with $\{0\} \times S \subset M$ and
$\Sigma$ with $\{0\} \times \partial S = S \cap {\cal J}$.
Let on ${\cal J} \equiv \mathbb{R} \times \partial S$  be given a smooth 3-dimensional  Lorentzian
conformal structure which satisfies in an adapted gauge together with the Cauchy data  the {\it corner conditions} implied on $\Sigma$  by the conformal field equations, where it is  assumed that the normals to $S$ are tangent to ${\cal J}$ on  $\Sigma$.

Then there exists for some $t_o > 0$ on the set  
\[
\hat{W} = \, ]- t_o,   t_o[ \times \hat{S} \subset \mathbb{R} \times \hat{S} \subset M
\]
a unique solution $\hat{g}$ to $Ric[\hat{g}] = \lambda\,\hat{g}$ which  admits 
with some smooth boundary defining function $\Omega$ on $M$ a smooth conformal extension 
\[
\hat{W}  \rightarrow W =   ]-   t_o,   t_o[ \times S,
\quad \hat{g}  \rightarrow g = \Omega^2\hat{g},
\]
that induces  (up to a conformal diffeomorphism) on $S$ and 
${\cal J}_o = ]-   t_o,   t_o[ \times \partial S$  the given conformal data.

\end{theorem}

It has been assumed here for convenience that all data are smooth and the corner conditions are satisfied at all orders. This ensures smoothness up to the boundary. If one is willing to accept some finite differentiability it should be observed that this might  entail a loss of differentiability at the boundary ${\cal J}$ (cf. \cite{benzoni-gavage:serre:2007} and the literature given there). Whether this loss does not occur because the fields satisfy besides the evolution equations also constraints has not been analysed.

In the following we give some background information, explain details, and point out particular features of AdS-type vacuum solutions which allow one to obtain this result.

\vspace{.2cm}

\noindent
{\it Existence of Cauchy data.}

\vspace{.2cm}

Concerning the construction of initial data it has been observed in \cite{friedrich:AdS} that there exists a correspondence between Cauchy data for the equations $Ric[\hat{g}] = \lambda\,\hat{g}$ with the appropriate behaviour at space-like infinity and  hyperboloidal data for the equations $Ric[\hat{g}] = 0$ which are conformally smooth at infinity. 
In fact, 
in the case of AdS-type solutions that are time reflection symmetric with respect to the initial slice $S$, so that the second fundamental form induced on it satisfies $\hat{\chi}_{ab} = 0$, the constraints reduce to
\[
R[\hat{h}] = 2\,\lambda = const. < 0,  
\]
and the solution must satisfy at space-like infinity an asymptotic behaviour  similar to that of hyperboloidal data. 
On the other hand,  there have been constructed in  \cite{andersson:chrusciel:friedrich:1992}
hyperboloidal data which are conformally smooth at infinity
under the assumption that $\lambda = 0$ and that the inner metric $\hat{h}_{ab}$ and the second fundamental form $\hat{\chi}_{ab}$ on the initial slice satisfy  $\hat{\chi}_{ab} =  \frac{\hat{\chi}}{3}\,\hat{h}_{ab}$, $\,\,\hat{\chi} = const. \neq 0$. In this case the constraints reduce to
\[
R[\hat{h}] = -  2\,\hat{\chi}^2/3 = const. < 0, 
\]
and it is seen  that the correspondence just requires a reinterpretation of the constants. In 
\cite{andersson:chrusciel:1996} the existence of a much more general class of conformally smooth hyperboloidal initial data has been shown and the generalization of the correspondence has been worked out in \cite{kannar:1996}.  

It may be mentioned here that in \cite{andersson:chrusciel:1996},  and even more so 
in \cite{andersson:chrusciel:friedrich:1992}, there have also been considered  hyperboloidal data
which only admit a poly-homogeneous expansion at infinity.
This may suggest to generalize the existence theorem cited above so as to include also boundary data on ${\cal J}$ with weaker smoothness requirements. The situation then becomes much more difficult, however, and special care must be taken to ensure that  no undesired non-smoothness is spreading into the space-time along the characteristic which generates the boundary of the domain of dependence of the data on $\hat{S}$.

\vspace{.2cm}

\noindent
{\it The boundary conditions/data.}

\vspace{.2cm}

Because no constraints are required on the conformal structure on ${\cal J}$ the prescription of the boundary data seems to be the easy part. There are, however, hidden subtleties here which are worth a detailed discussion. The initial boundary value problem  for the conformal field equations does not immediately lead to  the the conditions stated  in Theorem \ref{ads-exists}. To obtain the covariant formulation specific features of AdS-type solutions must be observed (cf. \cite{friedrich:geom-unique}   for a discussion of the problems which arise in other initial boundary value problems for Einstein's field equations).

The first of these special properties is the following. If $k_{ab}$ and $\kappa_{ab}$ denote the 
first  and second fundamental forms on ${\cal J}$ then the conformal field equations imply that the trace free part of $\kappa_{ab}$ vanishes so that
\begin{equation}
\label{kappa-vanishes}
\kappa_{ab} = 0 \quad \mbox{on} \quad  {\cal J},
\end{equation}
in a suitable conformal  gauge. This has the consequence that conformal geodesics (cf. \cite{friedrich:AdS}) which are tangent to ${\cal J}$ at one point stay in ${\cal J}$. 
Assuming for convenience that the space-like slice $S$ meets the boundary ${\cal J}$ orthogonally in the sense that the normal to $S$ is tangent to ${\cal J}$, this allows us  to set up the initial boundary value problem in terms of an {\it adapted conformal Gauss gauge}, which is generated by conformal geodesics that start orthogonally to $S$ with $g$-unit tangent vector. In particular, the coordinates are defined in terms of a natural parameter $\tau = x^0$ on the conformal geodesics, with $\tau = 0$  on $S$,
and coordinates $x^{\alpha}$ on $S$ which are extended so that they are constant along these curves.
The location of the boundary ${\cal J}$ is then determined  by the conformal geodesics which start on 
$\Sigma$ and the conformal factor $\Omega$, which vanishes on ${\cal J}$ and has non-vanishing differential there, is known explicitly  in terms of these coordinates.

In the adapted gauge the conformal field equations then imply a hyperbolic system of evolution equations which assumes in Newman-Penrose notation the form
\begin{equation}
\label{u-equ}
\partial_{\tau}u = F(u, \psi, x^{\mu}),
\end{equation}
\begin{equation}
\label{psi-equ}
(1 + A^0)\,\partial_{\tau}\psi + A^{\,\alpha}\,\,\partial_{\alpha}\psi = G(u, \psi, x^{\mu}).
\end{equation}
The unknown $u$ comprises the coefficients of a double null  frame field 
$(e^{\mu}\,_k )_{k = 0, .., 3} = (l^{\mu}, n^{\mu}, m^{\mu},  \bar{m}^{\mu})$ in these coordinates, the connection coefficients with respect to this frame, and the Schouten tensor 
\begin{equation}
\label{schouten}
L_{jk} = \frac{1}{n - 2}\,(R_{jk} - \frac{1}{2\,(n - 1)}\,R\,g_{jk}) 
\end{equation}
(with $n = 4$) of the metric $g$ in that frame. The matrices $A^{\mu}$ depend on the frame coefficients and the coordinates and $\psi = (\psi_0, \ldots \psi_4)$ 
represents the essential components of the symmetric spinor field $\psi_{ABCD}$ corresponding to the tensor field $W^{i}\,_{jkl}$.

To discuss the boundary conditions it is convenient to choose the frame such that the future directed time-like vector field $l + n$ is tangent to ${\cal J}$ and the space-like vector field 
$l - n$ is normal to ${\cal J}$ and inward pointing (this choice has in fact been included in the gauge conditions). Then $m$ and $\bar{m}$ are tangent to ${\cal J}$,
$l$ is inward  and $n$ is outward pointing on ${\cal J}$. This leads to boundary condition on ${\cal J}$ of the form
\begin{equation}
\label{geb-bdry-cond}
\psi_{4} - a\,\psi_{0} - c\,\bar{\psi}_{0} = d, \quad |a| + |c| \le 1, 
\end{equation}
where the smooth complex-valued function $d$ on ${\cal J}$ denotes the free boundary data and the smooth complex-valued functions $a$ and $c$  on ${\cal J}$ can be chosen freely within the indicated restrictions.

The form of the boundary conditions is made plausible by the following consideration.
It follows immediately from (\ref{u-equ}) that no part of $u$ can be prescribed. 
The Bianchi system $\nabla^A\,_{A'}\psi_{ABCD} = 0$, from which 
equation (\ref{psi-equ}) is extracted, splits into two subsystems. The first one, which is of the form
\[
\nabla_n \psi_k - \nabla_{m}\psi_{k + 1} = H_k(u, \psi) ,\quad k  = 0, 1, 2, 3,
\]
determines on ${\cal J}$ the outward transport of  $ \psi_0$, $\psi_1$, $ \psi_2$, $ \psi_3$ in terms of the fields  given in the interior and on the boundary. These components  of the rescaled Weyl tensor cannot be prescribed 
$ \psi_0$, $\psi_1$, $ \psi_2$, $ \psi_3$ on ${\cal J}$. The second subset, which is of the form
\[
\nabla_l \psi_j- \nabla_{\bar{m}}\psi_{j - 1} = K_j(u, \psi) ,\quad \,\,  j = 1, 2, 3, 4,
\]
describes an inward transport of the fields $\psi_1$, $ \psi_2$, $ \psi_3$, $ \psi_4$ on ${\cal J}$
and suggests that one may prescribe the field $\psi_4$ and possibly feed back into it some information on the other fields. The equations
\[
\nabla_{l+n} \psi_j- \nabla_{\bar{m}}\psi_{j - 1} - \nabla_{m}\psi_{j + 1} = 
H_j(u, \psi) + K_j(u, \psi) ,\quad \,\,  j = 1, 2, 3,
\]
implied by the equations above shows that the fields $\psi_1$, $ \psi_2$, $ \psi_3$ are governed by interior equations on ${\cal J}$. This implies with a detailed discussion involving energy estimates  that only information on $\psi_0$ can be fed back into $\psi_4$ and that the restriction given above has to be observed.

\vspace{.2cm}

\noindent
{\it The corner conditions}

\vspace{.2cm}

Given the conformal Cauchy data on $S$ and the adapted gauge in which the lines $x^{\alpha} = const.$ that start on ${\Sigma}$ are tangent to ${\cal J}$, the 
formal expansion of the unknowns in  equations (\ref{u-equ}), (\ref{psi-equ}) in terms of the coordinate  $\tau$ is determined at all orders uniquely by these equations on $S$ and  in particular on $\Sigma$. 
Let smooth functions $a$ and $c$ satisfying the restriction on the right hand side of 
(\ref{geb-bdry-cond}) be given on ${\cal J}$. Using their formal expansion in terms of $\tau$ 
on $\Sigma$, we get a formal expansion of the term on the right hand side of (\ref{geb-bdry-cond}) 
on $\Sigma$. The corner conditions consist in the requirement that this formal expansion 
coincides on $\Sigma$  with the formal expansion of the free boundary datum $d$.
Borel's theorem (\cite{dieudonne:I}) guarantees that there always exist smooth functions $d$ on ${\cal J}$ which satisfy this requirement. Away from $\Sigma$ they are essentially arbitrary.

\vspace{.2cm}

With Cauchy data as stated in the theorem and boundary conditions of the form
(\ref{geb-bdry-cond}) where $d$ satisfies the corner conditions one obtains a {\it well-posed initial boundary value problem which preserves the constraints and the gauge conditions}. This implies the  existence and uniqueness of smooth solutions on a domain as indicated in the theorem.

\vspace{.2cm}

\noindent
{\it Covariant boundary  conditions}

\vspace{.2cm}

The formulation obtained above has a drawback. Condition (\ref{geb-bdry-cond}) is not covariant, 
it depends in an implicit way on the choice of $l + n$. This is related to a general problem arising in initial boundary value problems for Einstein's field equations (cf.  \cite{friedrich:geom-unique}). 
In the case of  AdS-type solutions this problem can be overcome by making use of a second 
specific feature of such solutions.  They  always satisfy the relation
\begin{equation}
\label{w*-B-rel}
w^*_{ab}  = \sqrt{3/|\lambda|}\,B_{ab} \quad \mbox{on} \quad  {\cal J}.
\end{equation}
Here $w^*_{ab}$  denotes the ${\cal J}$-magnetic part of  $W^{i}\,_{jkl}$, obtained be contracting 
the right dual of $W^{i}\,_{jkl}$ twice with the inward pointing unit normal of ${\cal J}$, and 
$B_{ab}=  \epsilon_a\,^{cd}\,D_cL_{db}$  is the  Cotton tensor of the metric $k_{ab}$ on 
${\cal J}$, which is derived from the Schouten tensor  $L_{ab}$  of the metric $k_{ab}$ on ${\cal J}$ given by (\ref{schouten}) with $n = 3$ and $g$ replaced by $k$.

Two further observations are needed. The borderline case 
\begin{equation}
\label{magnetic-data}
\psi_{4} - \,\bar{\psi}_{0} = d
\end{equation}
of 
(\ref{geb-bdry-cond}) can be written in real notation with $d = d_1 + i\,d_2$  in the form
\[
w^*_{cd}\,M^{cd}\,_A = d_A , \quad A = 1, 2, 
\]
with constant real coefficients $M^{cd}\,_A$. Moreover, if the components $B_{cd}\,M^{cd}\,_A$ of the Cotton tensor are given on ${\cal J}$, the differential identity $D^aB_{ab} = 0$ of the Cotton tensor
turns into a hyperbolic system of PDE's for the remaining components if the frame vector fields  tangent to ${\cal J}$ and the connection defined by $k$ are known. 
The structural equations of the normal conformal Cartan connection of $k_{ab}$ allow us, however,  to deduce transport equations for the latter fields  so that  the combined system becomes hyperbolic.
From the Cauchy data on $S$ the field $B_{ab}$ and the lower order structures can be determined on $\Sigma$, which allows us integrate the conformal structure of $k_{ab}$ on ${\cal J}$. 
Here and below it is used that because of (\ref{kappa-vanishes})
the conformal geodesics with respect to $g$ which are lying in ${\cal J}$ are in fact also conformal geodesics with respect to the inner metric $k$ on ${\cal J}$.

It follows that the gauge underlying a boundary condition of the form (\ref{magnetic-data}) 
and the functions entering (\ref{magnetic-data}) can be determined on ${\cal J}$ in terms of the intrinsic conformal structure defined by $k$. 
PDE uniqueness then shows then that the `physical solution'
determined by the boundary condition (\ref{magnetic-data}) is determined uniquely by the Cauchy data and the conformal structure on ${\cal J}$.

\vspace{.2cm}

Any smooth AdS-type vacuum solution can be determined locally in time in the future of an initial slice like $S$ in terms of boundary conditions like (\ref{geb-bdry-cond}) or (\ref{magnetic-data}).  Besides leading to a covariant formulation, the conditions (\ref{magnetic-data}) are distinguished by another property.
Under the time reflection $\tau \rightarrow - \tau$ and the  transitions 
$\l \rightarrow - l$, $n \rightarrow - n$ the roles of $l$ and $n$
and of $\psi_4$ and $\psi_0$ 
in the discussion of the boundary conditions are swapped but the structure of the boundary condition (\ref{magnetic-data}) is preserved.
This is not necessarily true for other boundary  conditions given by (\ref{geb-bdry-cond}). It would be interesting to understand whether the freedom to choose the functions $a$ and $c$ on ${\cal J}$ has useful applications in the stability problem.

\section{Reflecting boundary condition}

Any of the boundary conditions  (\ref{geb-bdry-cond}) with $d = 0$ on ${\cal J}$ 
can be regarded as a reflecting boundary condition.
To ensure gauge independence we require 
\begin{equation}
\label{d-van-bdry-cond}
\psi_{4} - \,\bar{\psi}_{0} = 0 \quad \mbox{on} \quad {\cal J},
\end{equation}
and
\begin{equation}
\label{d-van-Sigma-bdry-cond}
\psi_{3} - \,\bar{\psi}_{1} = 0, \quad 
\psi_{2} - \,\bar{\psi}_{2} = 0\quad
\mbox{on} \quad \Sigma,
\end{equation}
which imply in particular that $w^*_{ab} = 0$ on $\Sigma$ and thus 
with  (\ref{w*-B-rel}) that 
\[
B_{ab}[k] = 0 \,\,\,\mbox{on}\,\,\, \Sigma.
\]
With the PDE implied by (\ref{d-van-bdry-cond}) and the identity $D^aB_{ab} = 0$ it follows then 
that (\ref{d-van-bdry-cond}), (\ref{d-van-Sigma-bdry-cond})  
are equivalent to  the {\it conformal flatness} condition 
\begin{equation}
\label{B-van-on-Sigma}
B_{ab}[k| = 0  \,\,\,\mbox{on}\,\,\, {\cal J}. 
\end{equation}
In the following we refer to this condition or the equivalent conditions (\ref{d-van-bdry-cond}) and
(\ref{d-van-Sigma-bdry-cond}) as the {\it reflecting boundary condition}.

Since (\ref{d-van-bdry-cond}) is by itself a `reflecting boundary condition' one may wonder why conditions 
(\ref{d-van-Sigma-bdry-cond}) are required as well. The reason, which is not immediately seen and will not be explained here in detail,
is that  (\ref{d-van-bdry-cond}) by itself still depends on the choice of the conformal gauge.

The analysis of the standard Cauchy problem admits a clean separation between the evolution problem and the construction of initial data. If  restrictions  are  imposed  on the boundary data in an initial boundary value problem this separation  cannot be maintained any longer. An extreme example of this situation is provided by reflecting boundary conditions. They do  not only prevent a flow of 
gravitational radiation across  ${\cal J}$, they also induce rather strong additional fall-off conditions on the initial data at  space-like infinity.

\begin{proposition}
\label{diff-conds}
In the situation considered in Theorem (\ref{ads-exists})
the reflecting boundary condition (\ref{B-van-on-Sigma}) imposes on the hyperboloidal Cauchy data set
$(S$, $\Omega$, $h_{ab}$ $\chi_{ab})$  and the rescaled Weyl spinor $\psi_{ABCD}$  calculated on $S$  from these data
not only the restrictions  $\psi_{4} = \bar{\psi}_{0}$, $\psi_{3} = \bar{\psi}_{1}$,  $\psi_{2} - \bar{\psi}_{2}$ 
at $\Sigma$,  but it implies with the corner conditions in addition a
sequence of differential conditions  on $\Sigma$ which involve derivatives of $\psi_{ABCD}$ of all orders.
\end{proposition}

\noindent
{\bf Proof}: The corner conditions imply with (\ref{d-van-bdry-cond}) the relations 
\begin{equation}
\label{corner-cond-inf-order}
\partial^k_{\tau}(\psi_{4} - \,\bar{\psi}_{0}) = 0, \,\,\,k = 0, 1, 2, \ldots
\quad \mbox{on} \quad \Sigma.
\end{equation}
With the evolution equations 
\[
(1 + A^0)\,\partial_{\tau}\psi + A^{\,\alpha}\,\,\partial_{\alpha}\psi = G(u, \psi),
\]
\[
\partial_{\tau}u = F(u, \psi, x^{\mu}),
\]
and their formal derivatives with respect to $\tau$ these conditions translate into inner differential conditions of all orders on the Cauchy data on $S$.\\
$\Box$

\vspace{.2cm}

If one is prepared to accept solutions of finite differentiability (and the associated drop of smoothness at the boundary), one could do with a finite number of derivatives in  
(\ref{corner-cond-inf-order}). The essential  problem in the construction of the Cauchy data still remains.

\section{Concluding remarks}

The requirement on the data in Proposition \ref{diff-conds}
raises the question to what extent data satisfying these conditions
can be provided in a systematic way by the known methods to construct hyperboloidal Cauchy data.
For simplicity we consider again the case treated in  \cite{andersson:chrusciel:friedrich:1992}. 
 It is shown there that, given a conformal structure on $\hat{S}$ in terms of a `seed metric` $\hat{h}$ which satisfies certain conditions at infinity, there exists a unique conformal factor $\Omega$ which satisfies together with the metric $h = \Omega^{-2}\,\hat{h}$ the conformal constraints and the required smoothness properties at infinity. Unless the differential conditions above can expressed for some unexpected reason directly in terms of conditions on the conformal structure defined by $\hat{h}$ it is hard to imagine that this method can be used to construct the data as required by Proposition. P. T. Chru\'sciel and E. Delay used hyperboloidal gluing techniques to show the existence of 
a class of non-trivial Cauchy data which are diffeomorphic to Schwarzschild-anti-de Sitter data outside some compact set and thus satisfy the corner conditions considered in Proposition \ref{diff-conds}
 (\cite{chrusciel:delay:2009}). To what extent this result can be generalized remains to be seen.

\vspace{.cm}

A second question is what the restriction on the data considered in Proposition \ref{diff-conds}
does to the development in time. Do the additional fall-off conditions increase the likeliness for the formation of trapped surfaces in the domain of dependence of the given Cauchy data ? In the case of asymptotically flat data there have been considered data
which also satisfy, besides the constraints and the asymptotically flatness condition, further conditions at space-like infinity (\cite{chrusciel:delay:2002}, \cite{corvino: 2007}). These do not give rise to problems in their development in time if the data were sufficiently small. 
In that case the additional  conditions have been imposed, however, 
only on the initial slice
while  everything else was left to the field equations and the solutions were only studied on the domain of dependence of the initial slice (and its conformal extension). An AdS-type solution satisfying reflective boundary conditions it required to develop as a curve in the set of restricted data as considered in  Proposition \ref{diff-conds} and, if possible, to develop beyond the domain of dependence of any space-like slice extending to ${\cal J}$.

\vspace{.2cm}

If the stability of solutions is to be analysed under less restrictive boundary conditions, there arises the question which boundary conditions/data should be given and what, in fact, should be meant by  `stability'.
As long as there are no physical considerations which could give a clue, the best one could do is perhaps 
to characterize in a most general way the  boundary conditions/data and Cauchy data for which solutions which start close to AdS stay close to AdS for all times. 
The global causal structure of AdS suggests that it will  in general not be reasonable to require 
anything more.

}


\begin{thebibliography}{9}


















\bibitem{andersson:chrusciel:1996}
L. Andersson, P. Chru\'sciel.
\newblock Solutions of the constraint equations in general relativity satisfying `hyperboloidal boundary conditions'.
\newblock {\em Dissertationes Mathematicae}, Polska Akademia Nauk, Inst. Matem.,
Warszawa, 1996.


\bibitem{andersson:chrusciel:friedrich:1992}
L. Andersson, P. Chru\'sciel, H. Friedrich.
\newblock On the existence of solutions to the Yamabe equation and the existence of smooth  hyperboloidal initial data for Einstein's field equations.
\newblock  {\em Commun. Math. Phys.} 149 (1992) 587 - 612.




\bibitem{avis:isham:storey:1978}
S. Avis, C. Isham, D. Storey.
\newblock Quantum field theory in anti-de Sitter space-time.
\newblock {\it Phys. Rev.} D 18 (1978) 3565 - 3576.


\bibitem{bachelot:2008}
A. Bachelot.
\newblock The Dirac system on the anti-de Sitter universe.
\newblock {\it Commun. Math. Phys.} 283 (2008) 127 - 167.

\bibitem{benzoni-gavage:serre:2007}
S. Benzoni-Gavage, D. Serre.
\newblock {\it Multi-dimensional hyperbolic partial differential equations}.
\newblock Clarendon Press, Oxford, 2007.

\bibitem{bizon:rostworowski:2011}
P. Bizo\'n, A. Rostworowski.
\newblock  On weakly turbulent stability of anti-de Sitter spacetime.
\newblock   {\em Phys. Rev. Lett.} 107 (2011) 031102. 

\bibitem{breitenlohner:freedman:1982}
P. Breitenlohner, D. Freedman.
\newblock Positive energy in anti-de Sitter backgrounds and gauge extended supergravity.
\newblock {\it Phys. Lett.} 115 B (1982) 197 - 201.

\bibitem{buchel-lehner-liebling:2012}
A. Buchel, L. Lehner, S. Liebling.
\newblock Scalar collapse in AdS spacetimes.
\newblock {\it Phys. Rev.} D 86 (2012) 123011.

\bibitem{buchel-lehner-liebling:2013}
A. Buchel, L. Lehner, S. Liebling.
\newblock Boson stars in AdS.
\newblock arXiv:1304.4166 [gr-qc]


\bibitem{chrusciel:delay:2002}
P. T. Chru\'sciel, E. Delay.
\newblock Existence of non-trivial, vacuum, asymptotically simple
spacetimes.
\newblock {\em Class. Quantum Grav.}, 19 (2002) L 71 - L 79.
\newblock Erratum
\newblock {\em Class. Quantum Grav.}, 19 (2002) 3389.


\bibitem{chrusciel:delay:2009}
P. T. Chru\'sciel, E. Delay.
\newblock Gluing constructions for asymptotically hyperbolic manifolds with constant scalar curvature.
\newblock {\it Communications in Analysis and Geometry} 17 (2009) 343 - 381.



\bibitem{corvino: 2007}
J. Corvino.
\newblock On the existence and stability of the Penrose compactification.
\newblock {\it Ann. Henri Poincar\'e} 8 (2007) 597 - 620.

\bibitem{dias:horowitz:santos:2012}
\'O. Dias, G. Horowitz, J. Santos.
\newblock Gravitational turbulent instability of anti-de Sitter space.
\newblock   {\it Class. Quantum Grav} 29 (2012) 194002. 

\bibitem{dias:horowitz:marolf:santos:2012}
\'O. Dias, G. Horowitz, D. Marolf, J. Santos.
\newblock On the non-linear stability of asymptotically anti-de Sitter solutions.
\newblock   {\it Class. Quantum Grav} 29 (2012) 235019. 

\bibitem{dieudonne:I}
J. Dieudonn\'e.
\newblock Foundations of moderns analysis.
\newblock Academic Press, New York, 1969.

\bibitem{fefferman:graham:1985}
C. Fefferman, R. Graham.
\newblock Conformal invariants.
\newblock In: {\it \'Elie Cartan et les math\'ematiques d'aujurd'hui} (Lyon 1984), {\it Ast\'erisque} (1985),
Num\'ero hors-s\'erie, 95 - 116.

\bibitem{friedrich:AdS}
H. Friedrich.
\newblock Einstein equations and conformal structure: existence of
anti-de Sitter-type space-times.
\newblock { \it J. Geom. Phys.}, 17 (1995) 125--184.

\bibitem{friedrich:geom-unique}
H. Friedrich.
\newblock Initial boundary value problem for Einstein's field equations and geometric uniqueness.
\newblock {\it Gen Relativ Gravit} 41 (2009) 1947 - 1966.



\bibitem{graham:hirachi:2005}
R. Graham, K. Hirachi.
\newblock The ambient obstruction tensor and Q-curvature.
In: O. Biquard (ed.) {\it AdS/CFT Correspondence: Einstein metrics and their conformal boundaries}.\newblock European Math. Soc., Z\"urich, 2005.







\bibitem{holzegel:2012}
G. Holzegel.
\newblock Well-posedness for the massive wave equation on asymptotically anti-de Sitter 
space-times.
\newblock {\it Journal of Hyperbolic Differential Equations} 9 (2012) 239 - 261.



\bibitem{ishibashi:maeda:2012}
A. Ishsibashi, K. Maeda.
\newblock Singularities in asymptotically anti-de Sitter space-times.
\newblock {\it Phys. Rev.} D 86 (2012) 104012




\bibitem{ishibashi:wald:2004} 
A. Ishibashi, R. Wald.
\newblock Dynamics in non-globally-hyperbolic static space-times: III. Anti-de Sitter space-time.
\newblock {\it Class. Quantum Grav.} 21 (2004) 2981 - 3013. 




\bibitem{kannar:1996}
J. K\'ann\'ar.
\newblock Hyperboloidal initial data for the vacuum Einstein equations with cosmological constant.
\newblock {\em Class. Quantum Grav.} 13 (1996) 3075 - 3084.

\bibitem{penrose:scri}
R. Penrose.
\newblock Zero rest-mass fields including gravitation: asymptotic behaviour.
\newblock {\em Proc. Roy. Soc. Lond.}, A 284 (1965) 159 - 203.

\bibitem{vasy:2010}
A. Vasy.
\newblock The wave equations on asymptotically anti-de Sitter spaces.
\newblock ADVAN MATH 11/2009; 223(1).


\bibitem{warnick:2012}
C. Warnick.
\newblock The massive wave equation in asymptotically AdS space-times. 
\newblock   arXiv:1202.3445 [gr-qc]





















\end{thebibliography}
\end{document}